\documentclass[twocolumn,reprint,aps,superscriptaddress,longbibliography]{revtex4-2}
\usepackage{amsmath}
\usepackage{amssymb}
\usepackage{amsfonts}
\usepackage[dvips]{graphicx}
\usepackage{subfigure}
\usepackage{dcolumn}
\usepackage{txfonts}
\usepackage{bm}
\usepackage{makeidx}
\usepackage{color}
\usepackage{mathtools}
\usepackage{threeparttable}
\usepackage[colorlinks,linkcolor=blue,anchorcolor=blue,citecolor=blue,urlcolor=blue]{hyperref}
\usepackage{lipsum}
\usepackage{braket}

\begin{document}

\title{Quantum imaging exploiting twisted photon pairs}

\author{Dianzhen Cui}

\author{X. X. Yi}
\email{yixx@nenu.edu.cn}

\author{Li-Ping Yang}
\email{lipingyang87@gmail.com}

\affiliation{Center for Quantum Sciences and School of Physics, Northeast Normal University, Changchun 130024, China}

\begin{abstract}
Quantum correlation of two-photon states has been utilized to suppress the environmental noise in imaging down to the single-photon level. However, the size of the coherence area of photon pairs limits the applications of quantum imaging based on spatial correlations. Here, we propose a quantum imaging scheme exploiting twisted photon pairs with tunable spatial-correlation regions to circumvent this limitation. We employ a bulk-density coincidence to enhance the imaging signal. Specifically, we introduce a re-scaled image signal, which is immune to the background intensity distribution profile of the photon pulse. We reveal a destructive interference between the anti-bunched photon pair and bunched photon pair in the imaging process. Our work could pave a way for twisted-photon-based quantum holography and quantum microscopy.
\end{abstract}

\maketitle

The quantum correlation of photon pairs in time and space offers a great advantage in quantum imaging~\cite{brida2010experimental,morris2015imaging,lemos2014quantum} and three-dimensional (3D) structure tomography~\cite{Nasr2003QOCT,Borja2020QOCT,lyons2018attosecond} down to the single-photon level. The correlated-photon imaging utilizing two-photon entanglement~\cite{Shih1995ghost} inspired streams of research in density-change-sensitive ghost imaging~\cite{bennink2002classical,Gatti2004thermal,valencia2005twophoton,cai2005ghost,Shapiro2008computational,Bromberg2009Ghost,Katz2009Compressive,Ferri2010differential} and phase-resolved quantum imgaing~\cite{Devaux2020imaging,Ndagano2022microscopy}. In addition to quantum imaging, the two-photon Hong-Ou-Mandel (HOM) interference, which is sensitive to the spatial phase-amplitude structure of input single photons, has also been exploited for the hologram of single photons~\cite{chrapkiewicz2016hologram} and high-dimensional photonic states engineering~\cite{zhang2016engineering}.

Spatial correlation is essential for correlated imaging. For a regular Gaussian photon pair from spontaneous parametric down-conversion (SPDC) processes, the size of its coherence area $\mathcal{A}_C=\pi R^2/k^2_0\sigma^2$~\cite{walborn2010spatial}, which is determined by the beam waist $\sigma$ of the pump beam, the center wave vector $k_0$, and the propagating distance $R$, limits the applications of quantum correlated imaging in many cases. A larger coherence area can be obtained by increasing the propagating distance $R$. However, this will attenuate the field strength significantly and reduce the signal-to-noise ratio (SNR). The tremendous advances in engineering complex optical fields open the possibility to develop photonic technologies for quantum imaging via precise manipulation of the transverse spatial properties of photons~\cite{Yu2011Light,Devlin2017,Shen2019,Padgett2017OAM,Loaiza2019iamging}.
Here we propose a quantum imaging scheme by exploiting twisted photon pairs with tunable coherence regions.

Twisted photons~\cite{Terriza2007Twisted} carrying quantized orbital angular momentum (OAM)~\cite{Allen1992OAM,yang2021quantum,Arnold2008OAM} stimulated much interest in high-dimensional quantum communication~\cite{Krenn2015twisted,wang2015Quantum,Agnew2011Tomography,shi2015Storage,Li2015Storage,Chen2021bright} and quantum computation~\cite{Babazadeh2017high,de2005implementing,Cozzolino2019high,krenn2014generation} beyond the polarization, momentum, and spectral degrees of freedom. Quantum imaging and remote sensing exploiting the continuous spatial correlation of twisted photon pairs will be another important topic of interest~\cite{Chen2014spiral,bornman2019ghost,Chen2019Quantum,tang2017spiral}. Usually, the field strength of a twisted photon is of a donut shape in the transverse plane. The radius of the donut (i.e., the coherence region of a twisted photon pair) can be tuned by varying the OAM quantum number. Thus, signal photons can be concentrated on the target regions to avoid SNR degradation. However, the non-uniform background intensity distribution profile of the photon pulse will hamper the imaging performance. How to resolve this difficulty remains elusive.

Our proposed quantum imaging is based on HOM interference with coincidence measurements between a bucket detector and an image sensor as shown in Fig.~\ref{fig:1}. Different from previous works sensing the intensity or phase change due to a quasi-transparent object~\cite{bornman2019ghost,Ndagano2022microscopy}, our imaging system aims to probe the texture by extracting the spatially varying phase imprinted on the photon during the reflection. The axial symmetry of twisted photons leads to completely destructive interference between the anti-bunched photon pair and bunched photon pair in the imaging process. This interference results in an effect that the texture information cannot be extracted via photon density measurements directly, but can be obtained from the quantum correlation of the two output photons. We also show that by retrieving the texture image via our introduced re-scaled signal, the influence of the photon density profile on quantum correlated imaging can be significantly eliminated. Our work constitutes valuable resources not only for quantum imaging or remote sensing but also for studying the unique quantum statistical properties of twisted photon pairs~\cite{yang2022quantum}.

\begin{figure*}
\centering
\includegraphics[width=18cm]{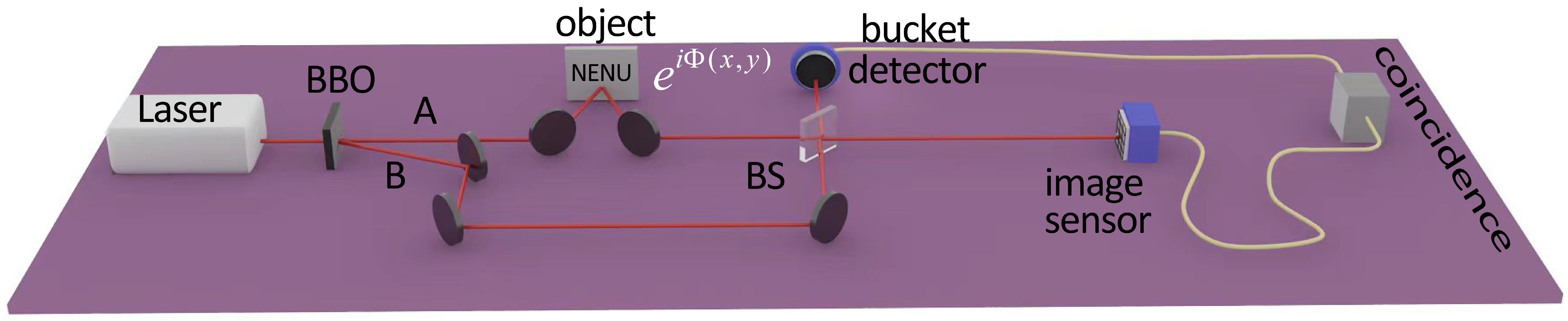}\caption{\label{fig:1} Schematic of the setup for our Hong-Ou-Mandel-based quantum imaging via twisted photon pairs. The two photons generated in the spontaneous down conversion processes propagate in different paths labeled by $A$ and $B$, respectively. The texture of the object is imprinted on the phase $\Phi(x,y)$ of the photon in path-$A$ during the reflection. Hong-Ou-Mandel interference of the two photons occurs at the beam splitter (BS). The information about the object is extracted via the coincidence of the photon number (obtained from the bucket detector) at one output port and the photon number density (measured via the image sensor) at the other port.}
\end{figure*}

\textit{Hong-Ou-Mandel interference of twisted photon pairs}\mbox{---}Our imaging scheme is based on HOM interferometer~\cite{HOM1987}. Here we give a general quantum theory of HOM interference, specifically for 3D structured photons. For two photons propagating in different directions, a linearly polarized photon pair can be described by a quantum state~\cite{supplementary}
\begin{equation}
\left|P_{\xi}\right\rangle =\int d\boldsymbol{k}\int d\boldsymbol{k}^{\prime}\xi(\boldsymbol{k}_{A},\boldsymbol{k}_{B}^{\prime})\hat{a}_{\boldsymbol{k}_{A}}^{\dagger}(t)\hat{b}_{\boldsymbol{k}_{B}^{\prime}}^{\dagger}(t)\left|0\right\rangle ,\label{eq:two-photon-state-1}
\end{equation}
where all the information of the photon pair is characterized by the spectral amplitude function (SAF) $\xi(\boldsymbol{k}_{A},\boldsymbol{k}_{B}^{\prime})$~\cite{loudon2000quantum} and two photonic paths are labeled by $A$ and $B$, respectively. Here a two-coordinate-frame formalism~\cite{walborn2003multimode} has been applied as shown in Fig.~\ref{fig:2} (a). For simplicity, we will only add the subscripts $A$ and $B$ in the quantum operators not in the SAF $\xi(\boldsymbol{k},\boldsymbol{k}^{\prime})$ in the following.  Different from the 
co-propagating two-photon state~\cite{yang2022quantum}, the spectral amplitude function (SAF) is not necessarily symmetrized~\cite{loudon2000quantum,supplementary}, i.e., $\xi (\boldsymbol{k},\boldsymbol{k}^{\prime})\neq \xi (\boldsymbol{k}^{\prime},\boldsymbol{k})$. In real space, the two-photon state can be rewritten as 
\begin{equation}
\left|P_{\xi}\right\rangle =\int d\boldsymbol{r}\int d\boldsymbol{r}^{\prime}\tilde{\xi}(\boldsymbol{r},\boldsymbol{r}^{\prime},t)\hat{\psi}_{a}^{\dagger}(\boldsymbol{r}_{A})\hat{\psi}_{b}^{\dagger}(\boldsymbol{r}_{B}^{\prime})\left|0\right\rangle, \label{eq:P_RS}   
\end{equation}
where $\hat{\psi}_{a(b)}(\boldsymbol{r})$ is the effective field operator of photons~\cite{Yang2021,supplementary} and the Fourier transformation of $\xi(\boldsymbol{k},\boldsymbol{k}')$ gives the wave-packet function (WPF)
\begin{equation}
\tilde{\xi}(\boldsymbol{r},\boldsymbol{r}^{\prime},t)=\frac{1}{(2\pi)^{3}}\int\!\! d\boldsymbol{k}\!\!\int\!\! d\boldsymbol{k}'\xi(\boldsymbol{k},\boldsymbol{k}^{\prime})e^{i(\boldsymbol{k}\cdot\boldsymbol{r}-\omega_{\boldsymbol{k}}t+\boldsymbol{k}^{\prime}\cdot\boldsymbol{r}^{\prime}-\omega_{\boldsymbol{k}'}t)}. \label{eq:xi_RS}
\end{equation}
We note that the WPF of each photon is also expressed in its co-moving frame~\cite{supplementary,walborn2003multimode}.

The HOM interference is essentially described by the input-output relations at the beam splitter [see Fig.~\ref{fig:2} (a)]~\cite{walborn2003multimode},
\begin{align}
\hat{c}_{\boldsymbol{k}_{B}} & =\left(R_{\boldsymbol{k}}\hat{a}_{\bar{\boldsymbol{k}}_{A}}+T_{\boldsymbol{k}}\hat{b}_{\boldsymbol{k}_{B}}\right),\\
\hat{d}_{\boldsymbol{k}_{A}} & =\left(T_{\boldsymbol{k}}\hat{a}_{\boldsymbol{k}_{A}}+R_{\boldsymbol{k}}\hat{b}_{\bar{\boldsymbol{k}}_{B}}\right),
\end{align}
with $\bar{\boldsymbol{k}}=(k_{x},-k_{y},k_{z})$. In the frame co-moving with photon, the $y$-component of wave vector changes sign under a reflection~\cite{supplementary,walborn2003multimode}. This leads to an important effect that the sign of the quantum number of photonic OAM is changed (i.e., $m\rightarrow -m$) under every reflection~\cite{Ritboon2019Optical,supplementary}. In the following, we focus on $50:50$ beam splitter with $T_{k}=1/\sqrt{2}$ and
$R_{k}=i/\sqrt{2}$ for paraxial quasi-single-frequency photons. Here, we only consider the HOM interference of two photons of the same polarization. Our formalism can be generalized to the cases for photons with different polarizations straightforwardly~\cite{branczyk2017hong,Ambrosio2019tunable}. 

The coincidence probability after the beam-splitter
\begin{align}
P_{{\rm cd}}^{(2)} & =\!\int\!\! d\boldsymbol{r}\!\!\int\!\! d\boldsymbol{r}'\left\langle \Psi_{\rm out}\right|\hat{\psi}_{c}^{\dagger}(\boldsymbol{r}_{B})\hat{\psi}_{d}^{\dagger}(\boldsymbol{r}_{A}^{\prime})\hat{\psi}_{d}(\boldsymbol{r}_{A}^{\prime})\hat{\psi}_{c}(\boldsymbol{r}_{B})\left|\Psi_{\rm out}\right\rangle,
\end{align} 
can be obtained from the output state~\cite{supplementary}
\begin{align}
&\left|\Psi_{{\rm out}}\right\rangle
=  \frac{1}{2}\int d\boldsymbol{r}\int d\boldsymbol{r}^{\prime}\left[i\tilde{\xi}(\boldsymbol{r},\bar{\boldsymbol{r}}^{\prime},t)\hat{\psi}_{d}^{\dagger}(\boldsymbol{r}_{A})\hat{\psi}_{d}^{\dagger}(\boldsymbol{r}_{A}^{\prime})\right. \nonumber \\
& \left.+i\tilde{\xi}(\bar{\boldsymbol{r}},\boldsymbol{r}^{\prime},t)\hat{\psi}_{c}^{\dagger}(\boldsymbol{r}_{B})\hat{\psi}_{c}^{\dagger}(\boldsymbol{r}_{B}^{\prime})+\tilde{\xi}_{cd}(\boldsymbol{r},\boldsymbol{r}^{\prime},t)\hat{\psi}_{d}^{\dagger}(\boldsymbol{r}_{A})\hat{\psi}_{c}^{\dagger}(\boldsymbol{r}_{B}^{\prime})\right]\left|0\right\rangle,\label{eq:Psi_out2} 
\end{align}
with $\bar{\boldsymbol{r}}=(x,-y,z)$ and
$\tilde{\xi}_{cd}(\boldsymbol{r},\boldsymbol{r}^{\prime},t)=\tilde{\xi}(\boldsymbol{r},\boldsymbol{r}^{\prime},t)-\tilde{\xi}(\bar{\boldsymbol{r}}^{\prime},\bar{\boldsymbol{r}},t)$. For quasi-1D photon pairs, the WPF is axially symmetric, i.e., $\tilde{\xi}(\bar{\boldsymbol{r}},\bar{\boldsymbol{r}}^{\prime},t)=\tilde{\xi}(\boldsymbol{r},\boldsymbol{r}^{\prime},t)$. Input two photons with an exchange-symmetric WPF $\tilde{\xi}(\boldsymbol{r},\boldsymbol{r}^{\prime},t)=\tilde{\xi}(\boldsymbol{r}^{\prime},\boldsymbol{r},t)$ will lead to vanishing  $\tilde{\xi}_{cd}(\boldsymbol{r},\boldsymbol{r}^{\prime},t)$ in the output state. Thus, two photons always come out from the same port due to destructive HOM interference. Input two photons with an exchange-antisymmetric WPF $\tilde{\xi}(\boldsymbol{r},\boldsymbol{r}^{\prime},t)=-\tilde{\xi}(\boldsymbol{r}^{\prime},\boldsymbol{r},t)$ 
lead to constructive HOM interference. Two photons always come out from different output ports resulting in a HOM peak~\cite{walborn2003multimode}. 

Different from the 1D case, the HOM interference of 3D twisted photon pairs becomes a little bit complicated. We now input an entangled twisted photon pair with WPF
\begin{equation}
\tilde{\xi}^{\pm}(\boldsymbol{r},\boldsymbol{r}^{\prime},t)=\mathcal{N}\tilde{\eta}(\boldsymbol{r},t)\tilde{\eta}(\boldsymbol{r}^{\prime},t)\left[e^{im(\varphi-\varphi^{\prime})}\pm e^{-im(\varphi-\varphi^{\prime})}\right], \label{eq:entangled_1}
\end{equation}
where the integer $m$ denotes the OAM quantum number of each photon, $\tilde{\eta}(\boldsymbol{r})$ characterizes the shape of each photon pulse and is usually independent on azimuthal angle $\varphi$, and $\mathcal{N}$ is the normalization factor. a HOM interference dip will be obtained for both the exchange-symmetric WPF $\tilde{\xi}^{+}(\boldsymbol{r},\boldsymbol{r}^{\prime},t)$ and the exchange-antisymmetric WPF $\tilde{\xi}^{-}(\boldsymbol{r},\boldsymbol{r}^{\prime},t)$ due to the fact $\tilde{\xi}^{\pm} (\bar{\boldsymbol{r}}',\bar{\boldsymbol{r}},t)=\tilde{\xi}^{\pm}(\boldsymbol{r},\boldsymbol{r}',t)$ (note the sign change in $m$ by $\bar{\boldsymbol{r}}$). For photon pairs with WPF
\begin{equation}
\tilde{\xi}(\boldsymbol{r},\boldsymbol{r}^{\prime},t)=\mathcal{N}\tilde{\eta}(\boldsymbol{r},t)\tilde{\eta}(\boldsymbol{r}^{\prime},t)\left[e^{im(\varphi+\varphi^{\prime})}-e^{-im(\varphi+\varphi^{\prime})}\right], \label{eq:entangled_2}
\end{equation}
a HOM interference peak will be obtained~\cite{zhang2016engineering}, since $\tilde{\xi} (\bar{\boldsymbol{r}}',\bar{\boldsymbol{r}},t)=-\tilde{\xi}(\boldsymbol{r},\boldsymbol{r}',t)$. More details can be found in the supplementary material~\cite{supplementary}.

\begin{figure}
\includegraphics[width=8cm]{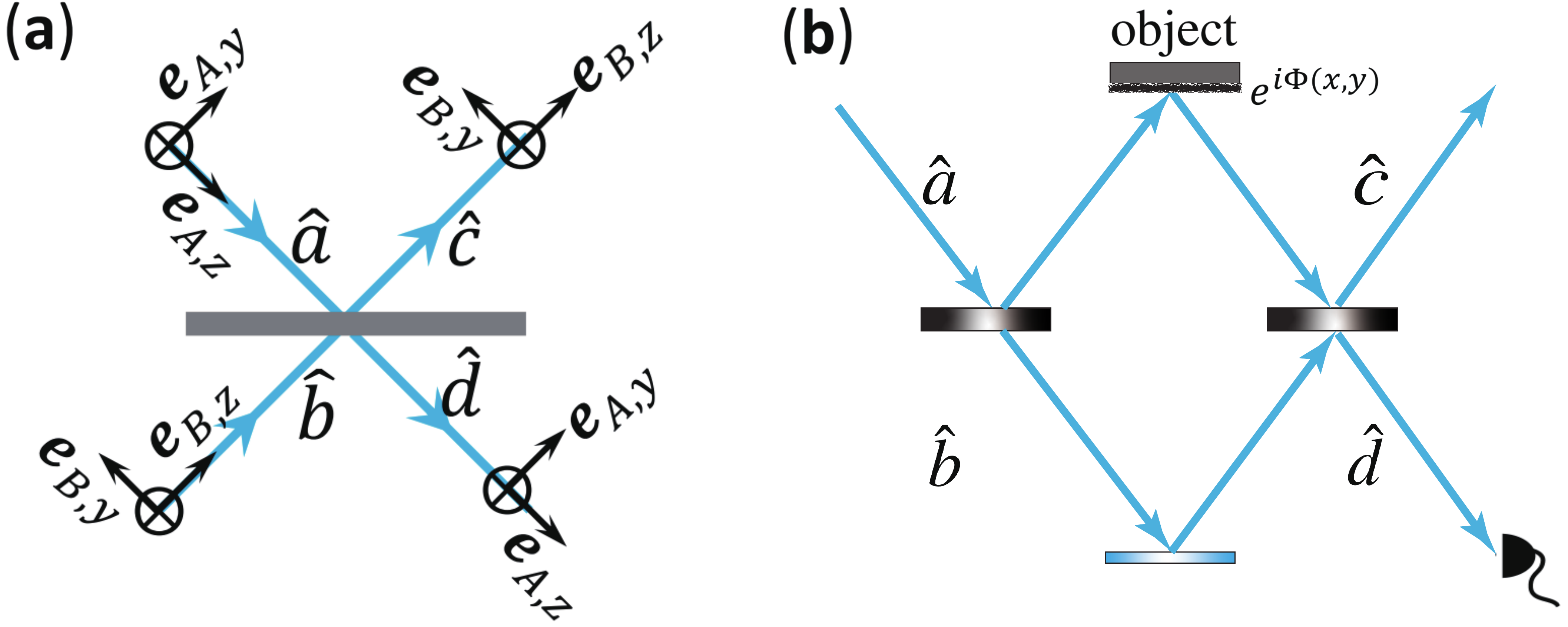}
\caption{\label{fig:2} (a) Transformation of the two coordinate frames at the beam splitter. These two coordinates corresponding to the two optical paths $A$ and $B$ are co-moving with the two photons. The input photonic modes are denoted by annihilation operators $\hat{a}$ and $\hat{b}$ and the output modes are denoted by $\hat{c}$ and $\hat{d}$. (b) Imaging via Mach-Zehnder interferometer with coherent-state input pulses. The texture of the object is imprinted on the phase $\Phi (x,y)$ in one optical channel and extracted via photon number density measurements directly. }
	
\end{figure}

\textit{HOM-Based quantum imaging}\mbox{---}Originally, the HOM interference was explored to measure the time delay (i.e., optical path difference) between the two incident photons~\cite{HOM1987}. Later on, it has been explored for various applications, such as quantum-optical coherence tomography~\cite{Nasr2003QOCT}, photon indistinguishability testing~\cite{santori2002indistinguishable}, quantum state engineering~\cite{lee2012second,zhang2016engineering}, as well as quantum imaging~\cite{chrapkiewicz2016hologram,Ndagano2022microscopy}. We now apply the HOM interferometer to detect the texture of an object with twisted photon pairs. As shown in Fig.~\ref{fig:1}, the photon in path-$A$ is reflected once by the object. The texture of the object is imprinted on the wave packet function of this photon by adding a spatially varying phase factor $\exp[i\Phi (x,y)]$ in the WPF $\tilde{\xi}(\boldsymbol{r},\boldsymbol{r}',t)$. After the HOM interference, this phase factor directly enters the WPF of output photons at both output ports as shown in Eq.~(\ref{eq:Psi_out2}).

In experiments, we can measure the photon number density at each output port, such as
\begin{equation}
n_{d}(\boldsymbol{r},t) =  \left\langle \Psi_{{\rm out}}\right|\hat{\psi}_{d}^{\dagger}(\boldsymbol{r}_{A})\hat{\psi}_{d}(\boldsymbol{r}_{A})\left|\Psi_{{\rm out}}\right\rangle, \label{eq:nd}
\end{equation} 
via a single-photon detector array, a CCD camera, or any other image sensor. Intuitively, we would expect to extract the phase factor $\Phi (x,y)$ directly from the photon-number density $n_{d}(\boldsymbol{r},t)$~\cite{lemos2014quantum}. However, this cannot be done in HOM-based imaging with twisted photon pairs as explained in the example. 

\begin{figure}
\centering
\includegraphics[width=8cm]{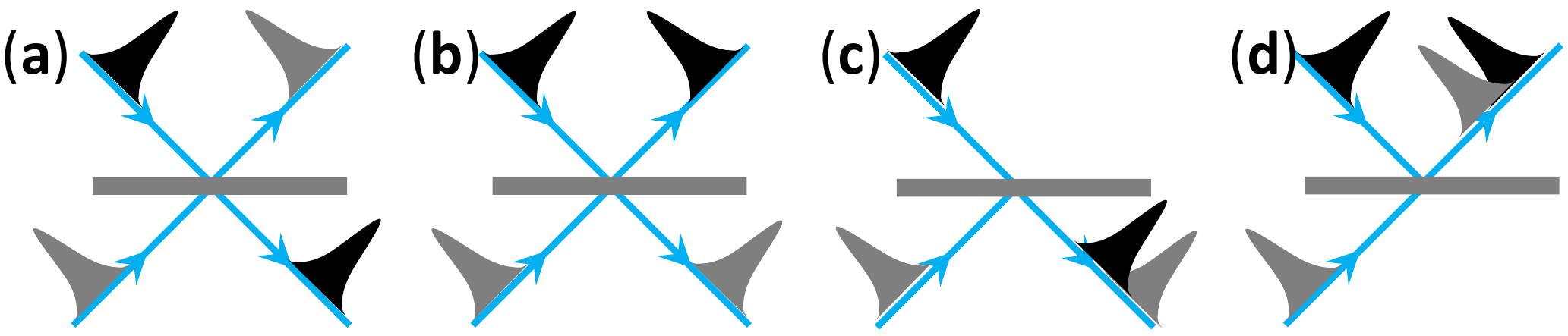}\caption{\label{fig:3} Four Hong-Ou-Mandel interference processes. (a) Both two photons are transmitted. (b) Both two photons are reflected. In (c) and (d), one photon is transmitted and the other is reflected. }
\end{figure}

The essence of the HOM interference lies in the beam-splitter-generated quantum entanglement between the two output photons. We now utilize the quantum correlation function $\left\langle \Psi_{{\rm out}}\right|\hat{\psi}_{d}^{\dagger}(\boldsymbol{r}_{A})\hat{\psi}_{c}^{\dagger}(\boldsymbol{r}_{B}^{\prime})\hat{\psi}_{c}(\boldsymbol{r}_{B}^{\prime})\hat{\psi}_{d}(\boldsymbol{r}_{A})\left|\Psi_{{\rm out}}\right\rangle$ to extract the 
spatially dependent phase $\Phi(x,y)$. To obtain a larger signal and to speed up the imaging process, we perform the following coincidence signal detection
\begin{equation}
\!\!\!\!\langle\hat{\mathcal{C}}_{d}(\boldsymbol{r},t)\rangle  =\!\!\int\!\! d\boldsymbol{r}'\left\langle \Psi_{{\rm out}}\right|\hat{\psi}_{d}^{\dagger}(\boldsymbol{r}_{A})\hat{\psi}_{c}^{\dagger}(\boldsymbol{r}_{B}^{\prime})\hat{\psi}_{c}(\boldsymbol{r}_{B}^{\prime})\hat{\psi}_{d}(\boldsymbol{r}_{A})\left|\Psi_{{\rm out}}\right\rangle.  
\end{equation}
As shown in Fig.~\ref{fig:1}, a bucket detector is employed to collect the photon number $\int d\boldsymbol{r}\langle\hat{\psi}^{\dagger}_c(\boldsymbol{r}_B)\hat{\psi}_c(\boldsymbol{r}_B)\rangle$ at port-$c$ and an image sensor is used to measure the photon number density $n_d(\boldsymbol{r},t)$ at port-$d$. The quantum imaging of the texture of the object is achieved via the coincidence signal $\langle\hat{\mathcal{C}}_{d}(\boldsymbol{r},t)\rangle$.

In practice, we cannot measure the true photon number density at a single point. Instead, we measure the accumulated signal at a finite small volume $\Delta V$ determined by the pixel area and measuring time. Thus, the measured signal from the pixel labeled by $\boldsymbol{X}$ will be the mean value of the operator
\begin{align}
\hat{\mathcal{C}}_{d}(\boldsymbol{X}) & =\int_{\Delta V_{\boldsymbol{X}}} d\boldsymbol{r}\int d\boldsymbol{r}'\hat{\psi}_{d}^{\dagger}(\boldsymbol{r}_{A})\hat{\psi}_{c}^{\dagger}(\boldsymbol{r}_{B}^{\prime})\hat{\psi}_{c}(\boldsymbol{r}_{B}^{\prime})\hat{\psi}_{d}(\boldsymbol{r}_{A}),\label{eq:C_pixel}
\end{align}
where the integral over $\boldsymbol{r}$ is limited in the small volume $\Delta V_{\boldsymbol{X}}$ corresponding to the pixel-$\boldsymbol{X}$. This also removes the divergence in the density correlation at the same point $\langle\Psi_{\rm out}|[\hat{\psi}^{\dagger}_d(\boldsymbol{r}_A)\hat{\psi}_d(\boldsymbol{r}_A)]^2|\Psi_{\rm out}\rangle$~\cite{yang2021quantum}. The SNR for two-photon-state imaging after $N$ independent measurements
is defined as~\cite{Ferri2010differential,supplementary}
\begin{align}
{\rm SNR}_{\rm TPS} (\boldsymbol{X}) & =\frac{\sqrt{N}\langle\hat{\mathcal{C}}_d(\boldsymbol{X})\rangle}{\sqrt{\langle\Delta\hat{\mathcal{C}}^2_d(\boldsymbol{X})\rangle}}=\frac{\sqrt{N}\langle\hat{\mathcal{C}}_d(\boldsymbol{X})\rangle}{\sqrt{\langle\hat{\mathcal{C}}_{d}(\boldsymbol{X})\rangle- \langle\hat{\mathcal{C}}_{d}(\boldsymbol{X})\rangle^2}},
\end{align}
where $\langle\hat{\mathcal{C}}_d(\boldsymbol{X})\rangle\ll 1$, since it characterizes the probability of detecting a photon by pixel-$\boldsymbol{X}$.

The object texture can also be extracted via the Mach-Zehnder interference with coherent-state laser pulses [see Fig.~\ref{fig:2} (b)]. The image signal is obtained by the pixelated photon number density operator $\hat{\mathcal{D}}_d (\boldsymbol{X})=\int_{\Delta V_{\boldsymbol{X}}} d\boldsymbol{r}\hat{\psi}_{d}^{\dagger}(\boldsymbol{r}_{A})\hat{\psi}_{d}(\boldsymbol{r}_{A})$. The SNR for coherent-state imaging is
\begin{align}
{\rm SNR}_{\rm CS} (\boldsymbol{X}) & =\frac{\langle\hat{\mathcal{D}}_d(\boldsymbol{X})\rangle}{\sqrt{\langle\Delta\hat{\mathcal{D}}^2_d(\boldsymbol{X})\rangle}}=\frac{\langle\hat{\mathcal{D}}_d(\boldsymbol{X})\rangle}{\sqrt{\langle\hat{\mathcal{D}}_{d}(\boldsymbol{X})\rangle}}.
\end{align}
For a coherent-state pulse with a large photon number $N$, the well-known $\sqrt{N}$-factor enhancement in ${\rm SNR}_{\rm CS} (\boldsymbol{X})$ will be obtained. Comparing ${\rm SNR}_{\rm TPS} (\boldsymbol{X})$ and ${\rm SNR}_{\rm CS} (\boldsymbol{X})$, we see that the quantum imaging based on two-photon HOM interference enhance the SNR slightly at the single-photon level. However, due to  technological limitations, the dark-counting related noise dominates in quantum correlated imaging instead of the quantum noise in ${\rm SNR}_{\rm TPS} (\boldsymbol{X})$~\cite{moreau2019imaging}. In the following, we focus more on the theoretical exploration of exploiting twisted photon pairs in quantum imaging. Additionally, we show twisted photon pairs can be used for the quantum encryption of images via HOM interference.

\textit{Quantum imaging with twisted photon pairs}---We now apply our quantum imaging approach to specific twisted photon pairs. The input two photons in a product state are described by Eq.~(\ref{eq:P_RS}) with WPF
\begin{equation}
\tilde{\xi}(\boldsymbol{r},\boldsymbol{r}^{\prime},t)=\mathcal{N}\tilde{\eta}_{m}(\rho,z,t)\tilde{\eta}_{-m}(\rho^{\prime},z',t)e^{i[m(\varphi-\varphi^{\prime})+\Phi(\rho,\varphi)]}. \label{eq:WPF-1}
\end{equation}
Here, the two photons have opposite OAM quantum numbers. Only the photon in path-$A$ has been reflected by the object. Thus, the texture information of the object is imprinted on the WPF of this photon via the phase $\Phi(\rho,\varphi)$ re-expressed in a cylindrical coordinate. Even for a product input state, the HOM interferometer generates quantum entanglement in the output photons, which plays an essential role in the imaging process.

The photon number density (\ref{eq:nd}) measured from output port-$d$ is given by
\begin{equation}
n_{d}(\boldsymbol{r},t) = \frac{1}{4}\left|\tilde{\eta}_{m}(\rho,z,t)\right|^{2}\left\{ 4+\left[(I_{1}-I_{2})e^{i\Phi(\rho,\varphi)}+{\rm c.c.}\right]\right\},  
\end{equation}
where the overlap integrals $I_{1} =\int d\boldsymbol{r}'\left|\tilde{\eta}_{m}(\rho^{\prime},z',t)\right|^{2}e^{-i\Phi(\rho^{\prime},\varphi^{\prime})}$ and $I_{2}   =\int d\boldsymbol{r}'\left|\tilde{\eta}_{m}(\rho^{\prime},z',t)\right|^{2}e^{-i\Phi(\rho^{\prime},-\varphi^{\prime})}$
come from the first term (bunched photon pair) and third term (anti-bunched photon pair) of the output state (\ref{eq:Psi_out2}), respectively. For an axisymmetric function $\tilde{\eta}_m(\rho,z,t)$, we can prove $I_1=I_2$~\cite{supplementary}, thus the corresponding terms carrying texture information in $n_d(\boldsymbol{r},t)$ cancel out. Interference between the anti-bunched pair [processes (a) and (b) in Fig.~\ref{fig:3}] and bunched pair [process (c) in Fig.~\ref{fig:3}] occurs. This destructive interference leads to a striking effect that the texture of the object cannot be extracted via simply measuring $n_d(\boldsymbol{r},t)$. We note that this completely destructive interference is essentially due to the axial symmetry of twisted photons and it is significantly different from the well-known interference resulting in the HOM dip or peak in $P^{(2)}_{cd}$, which occurs only between the anti-bunched photons [the process (a) and the process (b) in Fig.~\ref{fig:3}]. The numerical simulation of $n_d(\boldsymbol{r},t)$ is shown by the top row in Fig.~\ref{fig:4}. Only the donut-structure of twisted light has been observed. This also shows a fundamental departure from the interference of two coherent-state pulses in a Mach-Zehnder experiment, where the photon number density at the output port will be $\propto |\exp[im\varphi+i\Phi(x,y)]+\exp(im\varphi)|^2$. 

\begin{figure}
\centering
\includegraphics[width=8cm]{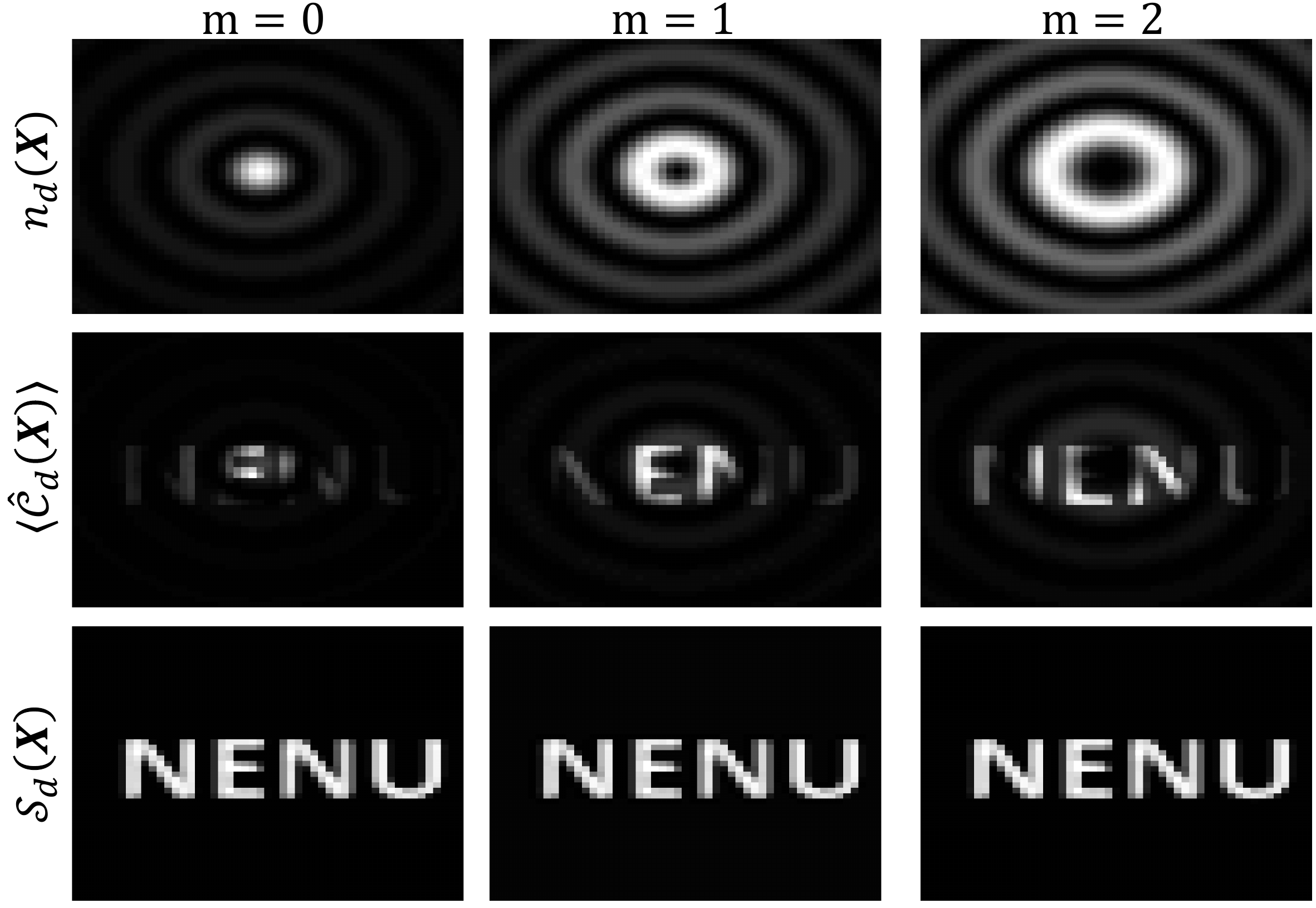}\caption{\label{fig:4} Photon number density $n_d(\boldsymbol{X})$ (the top row), coincidence signal $\langle\hat{\mathcal{C}}_d(\boldsymbol{X})\rangle$ (middle row), and the rescaled signal $\mathcal{S}_d(\boldsymbol{X})$ (bottom row) for product-state photon pairs described by Eq.~(\ref{eq:WPF-1}). Different columns correspond to twisted photon pairs with different orbital angular momentum quantum numbers $m$. Here, the accumulated signal on a pixel [see Eq.~({\ref{eq:C_pixel}})] not the density at a single point has been shown. More details about the numerical simulation can be found in supplementary material~\cite{supplementary}.}
\end{figure}

The coincidence signal is given by
\begin{equation}
\langle\hat{\mathcal{C}}_d (\boldsymbol{r},t)\rangle =\frac{1}{4}\left|\tilde{\eta}_{m}(\boldsymbol{r},t)\right|^{2}\left[2-\left(I_{2}e^{i\Phi(\rho,\varphi)}+\rm{c.c.}\right)\right].
\end{equation}
During the calibration by replacing the object with a mirror (i.e., $\Phi=0$), the optical paths of the two photons have been carefully matched and no coincidence signal will be obtained $P^{(2)}_{\rm cd} =0$. The texture of the object introduces extra phase difference $\Phi(\rho,\varphi)$ resulting in non-vanishing coincidence signal $\langle\hat{\mathcal{C}}_d (\boldsymbol{r},t)\rangle$. As shown by the middle row of Fig.~\ref{fig:4}, the texture of the object looms up in $\langle\hat{\mathcal{C}}_d (\boldsymbol{r},t)\rangle$. By varying the OAM quantum number $m$, the imaging region can be tuned gradually.

To remove the influence of the background density distribution profile of twisted photons, we introduce a re-scaled signal
\begin{equation}
\mathcal{S}_{d}(\boldsymbol{r},t)\!\equiv\!\frac{\langle\hat{\mathcal{C}}_d (\boldsymbol{r},t)\rangle\!-\!n_{d}(\boldsymbol{r},t)/2}{n_{d}(\boldsymbol{r},t)/2}\!=\!-\frac{1}{2}\left[I_{2}e^{i\Phi(\rho,\varphi)}\!+\!{\rm c.c.}\right].
\end{equation} 
As shown by the bottom row of Fig.~\ref{fig:4}, a texture image of the object with a much higher contrast is obtained. Due to the perfect destructive interference, no texture information of the object is contained in the photon number density $n_d(\boldsymbol{r},t)$. The imaging signal has been completely extracted via $\mathcal{S}_d(\boldsymbol{r},t)$. In this work, only the quantum imaging with product-state photon pairs has been investigated. Our approach can be directly applied to the entangled twisted photon pairs, such as the two-photon states in Eqs.~(\ref{eq:entangled_1}) and (\ref{eq:entangled_2}). Helical-phase-modified quantum coherence of entangled twisted photon pairs~\cite{yang2022quantum} will manifest in the texture image as shown in our following work.

Our discovered destructive interference effect in the imaging process can be exploited for the quantum encryption of images with twisted photon pairs. Similar to our imaging process, we first encode an image in one of the photon's phase and then perform a HOM interference. Based on our results, the information of the photo cannot be extracted via the photon number density at either output. Retrieval of the photo can only be achieved via the coincidence signal. In our theoretical simulation, we only take a twisted pair with WPF in a product form for simplicity. Two photons generated in an SPDC process are entangled in frequency degrees of freedom~\cite{Hong1985SPDC,Monken1998SPDC,walborn2010spatial}. However, the frequency entanglement does not change the main features of a HOM interference~\cite{branczyk2017hong}. In experiments, spectral filters can be used to ensure spectral indistinguishability of the two photons and to remove the frequency entanglement~\cite{Scott2017limits,Lyons2018Ashley}.

\textit{Discussion}.\mbox{---}We conduct a theoretical exploration of quantum imaging with twisted photon pairs in this letter. In our numerical simulation, diffraction-free Bessel pulses with micrometer-scale cross-sections have been taken. However, our quantum imaging approach is not limited to microscopy. Laguerre-Gaussian-mode twisted photons can be used for imaging a macroscopic object. The advances in single-photon-level image sensors~\cite{morimoto2020megapixel,piron2020review} lay a solid hardware foundation for our proposed experiment. To suppress the imaging noise in experiments, we can replace the bucket detector with a superconducting nanowire single-photon detector with an extremely low dark-counting rate~\cite{schuck2013waveguide}. The new generation of Megapixel single-photon avalanche photodiode image sensors with smaller pitch size ($<10\ {\rm \mu m}$), higher frame rate and time resolution will enable more exciting applications of quantum imaging at single-photon level~\cite{morimoto2020megapixel,shin2016photon}. 

\textit{Acknowledgements}.\mbox{---}The authors thank H. Dong, D. Z. Xu, and S. W. Li for fruitful discussions. This work is supported by National Key R\&D Program of China (Grant No. 2021YFE0193500), NSFC Grant
No.12275048, and NSFC Grants No. 12175033.

\bibliography{main}


\widetext
\newpage
\begin{center}
\textbf{\large Supplementary Materials for ``Quantum imaging exploiting twisted photon pairs''}
\end{center}

\maketitle

\tableofcontents

In this supplementary, we give the details in the evaluation of  photon number densities and photon number density correlation functions in the Hong-Ou-Mandel (HOM) interference with twisted photon pairs.    

\maketitle
\begin{appendix}

\section{Two-coordinate-frame description of photon pairs}
In this section, we give a detailed introduction to the two-coordinate-frame formalism, which provides a convenient way to handle two-channel interference processes, such as the Mach-Zehnder interference and Hong-Ou-Mandel (HOM) interference. The quantum state of a photon pair can be generally expanded with plane-wave modes
\begin{equation}
\left|P_{\xi}\right\rangle =\frac{1}{\sqrt{2}}\int d\boldsymbol{k}\int d\boldsymbol{k}'\xi(\boldsymbol{k},\boldsymbol{k}')\hat{a}_{\boldsymbol{k}}^{\dagger}\hat{a}_{\boldsymbol{k}'}^{\dagger}\left|0\right\rangle,\label{eq:two-photon-state}
\end{equation}
where $\xi(\boldsymbol{k},\boldsymbol{k}')$ is the spectral-amplitude function (SAF) and we do not consider the polarization degrees of freedom in this work. Due to the bosonic commutations relation $[\hat{a}_{\boldsymbol{k}}^{\dagger},\hat{a}_{\boldsymbol{k}'}^{\dagger}]=0$,
the quantum state $\left|P_{\xi}\right\rangle $ does not change under
the permutation $\boldsymbol{k}\rightarrow\boldsymbol{k}'$. Thus,
the two-photon is required to be
symmetric 
\begin{equation}
\xi(\boldsymbol{k},\boldsymbol{k}')=\xi(\boldsymbol{k}',\boldsymbol{k}).
\end{equation}
The normalization constraint for the SAF $\int d^{3}k\int d^{3}k'\left|\xi(\boldsymbol{k},\boldsymbol{k}')\right|^{2}=1$ is obtained via $[\hat{a}_{\boldsymbol{k}},\hat{a}_{\boldsymbol{k}'}^{\dagger}]=\delta(\boldsymbol{k}-\boldsymbol{k}')$. 

We note that the plane-wave expansion of the quantum state of a photon pair in (\ref{eq:two-photon-state}) has
been performed in the same coordinate frame. However, for two photons
propagating in two different directions, the SAF $\xi(\boldsymbol{k},\boldsymbol{k}')$
will become extremely complicated usually. For paraxial photon
pulses, the probability of two photons having the same wave vector $\boldsymbol{k}$
will be negligibly small, $|\xi(\boldsymbol{k},\boldsymbol{k})|^{2}\rightarrow0$.
Thus, it will be much more convenient to introduce the two-coordinate-frame formalism~\cite{walborn2003multimode}. In this case, the two-photon
a quantum state can be approximated as
\begin{equation}
\left|P_{\xi}\right\rangle \approx\int d\boldsymbol{k}\int d\boldsymbol{k}^{\prime}\xi(\boldsymbol{k}_A,\boldsymbol{k}_B^{\prime})\hat{a}_{\boldsymbol{k}_{A}}^{\dagger}\hat{b}_{\boldsymbol{k}_{B}^{\prime}}^{\dagger}\left|0\right\rangle ,\label{eq:two-photon-state-1}
\end{equation}
the two photonic paths are labeled by $A$ and $B$, respectively.
For each path, we establish a separate coordinate co-moving with the photon as shown in Fig.~\ref{fig:beamsplitter} (a), i.e., $\boldsymbol{k}_{A}=k_{x}\boldsymbol{e}_{A,x}+k_{y}\boldsymbol{e}_{A,y}+k_{z}\boldsymbol{e}_{A,z}$
and $\boldsymbol{k}_{B}=k_{x}\boldsymbol{e}_{B,x}+k_{y}\boldsymbol{e}_{B,y}+k_{z}\boldsymbol{e}_{B,z}$. We also use different ladder operators to denote the two photons,
because the corresponding ladder operators commutes, i.e., $[\hat{a}_{\boldsymbol{k}_{A}},\hat{b}_{\boldsymbol{k}_{B}^{\prime}}^{\dagger}]=[\hat{b}_{\boldsymbol{k}_{B}},\hat{a}_{\boldsymbol{k}_{A}^{\prime}}^{\dagger}]=0$. For simplicity, we only add the subscripts $A$ and $B$ to the wave vectors in the ladder operators not in the SAF in the following. As shown in Sec.~\ref{sec:beamsplitter}, the coupling between plane wave modes at a beam splitter can be well described by the indices of the ladder operators. The SAF $\xi(\boldsymbol{k},\boldsymbol{k}^{\prime})$ still satisfies the normalization constraint $\int d\boldsymbol{k}\int d\boldsymbol{k}^{\prime}\left|\xi(\boldsymbol{k},\boldsymbol{k}^{\prime})\right|^{2}=1$. 

Usually, the SAF $\xi(\boldsymbol{k},\boldsymbol{k}^{\prime})$
does not satisfy the symmetry requirement $\xi(\boldsymbol{k},\boldsymbol{k}^{\prime})=\xi(\boldsymbol{k}^{\prime},\boldsymbol{k})$, because these two photons can be distinguished via their propagating paths. As pointed
out by Leggett~\cite{leggett2006quantum}: ``It is crucial to appreciate
that the mere fact that a given system of particles shows substantial
effects of quantum mechanics such as the quantization of energy are
not enough to guarantee that it will automatically show the effects
of indistinguishability; it is necessary, in addition, for the particles
to be able to \textquotedblleft find out\textquotedblright{} that
they are indistinguishable, and they can do this only if they can
change places (otherwise, we can as it were \textquotedblleft tag\textquotedblright{}
them by their physical location).'' Only when the two photons meet at the beam splitter, the effect of indistinguishability manifests again that we can not tell the output photon from which input port. This indistinguishability plays an essential role in the quantum description of a beam splitter as shown in the following.

In the Schr\"odinger picture, the state of the photon pair $\left|P_{\xi}(t)\right\rangle $
at time $t$ can be obtained by replacing the ladder operators with
\begin{align}
e^{-i\hat{H}t/\hbar}\hat{a}_{\boldsymbol{k}_{A}}^{\dagger}e^{i\hat{H}t/\hbar} & =\hat{a}_{\boldsymbol{k}_{A}}^{\dagger}e^{-i\omega_{\boldsymbol{k}}t},\\
e^{-i\hat{H}t/\hbar}\hat{b}_{\boldsymbol{k}_{B}}^{\dagger}e^{i\hat{H}t/\hbar} & =\hat{b}_{\boldsymbol{k}_{B}}^{\dagger}e^{-i\omega_{\boldsymbol{k}}t}.
\end{align}
We now introduce two effective field operators in the real-space
\begin{align}
\hat{\psi}_{a}(\boldsymbol{r}_{A}) & =\frac{1}{\sqrt{(2\pi)^{3}}}\int d\boldsymbol{k}\hat{a}_{\boldsymbol{k}_{A}}e^{i\boldsymbol{k}_A\cdot\boldsymbol{r}_{A}},\\
\hat{\psi}_{b}(\boldsymbol{r}_{B}) & =\frac{1}{\sqrt{(2\pi)^{3}}}\int d\boldsymbol{k}\hat{b}_{\boldsymbol{k}_{B}}e^{i\boldsymbol{k}_{B}\cdot\boldsymbol{r}_{B}},
\end{align}
where $\boldsymbol{r}_{A}=x\boldsymbol{e}_{A,x}+y\boldsymbol{e}_{A,y}+z\boldsymbol{e}_{A,z}$ and
$\boldsymbol{r}_{B}=x\boldsymbol{e}_{B,x}+y\boldsymbol{e}_{B,y}+z\boldsymbol{e}_{B,z}$.
We rewrite the quantum state of a photon pair as
\begin{align}
\left|P_{\xi}(t)\right\rangle & =\int d\boldsymbol{r}\int d\boldsymbol{r}^{\prime}\hat{\psi}_{a}^{\dagger}(\boldsymbol{r}_{A})\hat{\psi}_{b}^{\dagger}(\boldsymbol{r}_{B}^{\prime})\frac{1}{(2\pi)^{3}}\int d\boldsymbol{k}\int d\boldsymbol{k}^{\prime}\xi(\boldsymbol{k}_{A},\boldsymbol{k}_{B}^{\prime})e^{i(\boldsymbol{k}_{A}\cdot \boldsymbol{r}_{A}-\omega_{\boldsymbol{k}}t+\boldsymbol{k}_{B}^{\prime}\cdot\boldsymbol{r}_{B}^{\prime}-\omega_{\boldsymbol{k}'}t)}\left|0\right\rangle \\
 & =\int d\boldsymbol{r}\int d\boldsymbol{r}^{\prime}\tilde{\xi}(\boldsymbol{r},\boldsymbol{r}^{\prime},t)\hat{\psi}_{a}^{\dagger}(\boldsymbol{r}_{A})\hat{\psi}_{b}^{\dagger}(\boldsymbol{r}_{B}^{\prime})\left|0\right\rangle, \label{eq:P_RS}
\end{align}
with the wave-packet function of the photon pair
\begin{equation}
\tilde{\xi}(\boldsymbol{r},\boldsymbol{r}^{\prime},t)=\frac{1}{(2\pi)^{3}}\int d\boldsymbol{k}\int d\boldsymbol{k}'\xi(\boldsymbol{k},\boldsymbol{k}^{\prime})e^{i(\boldsymbol{k}\cdot\boldsymbol{r}-\omega_{\boldsymbol{k}}t+\boldsymbol{k}^{\prime}\cdot\boldsymbol{r}^{\prime}-\omega_{\boldsymbol{k}'}t)}. \label{eq:xi_RS}
\end{equation}
We note that this two-coordinate-frame formalism only works for paraxial pulses, in which photons can be well distinguished via their propagating axes.

\subsection*{Examples}

The SAF of a photon pair generated via spontaneous parametric down-conversion (SPDC) processes is given by~\cite{Hong1985SPDC,walborn2003multimode,Monken1998SPDC}
\begin{equation}
\xi(\boldsymbol{k},\boldsymbol{k}^\prime)=\frac{1}{\pi}\sqrt{\frac{2L}{K}}\eta(\boldsymbol{k}+\boldsymbol{k}^\prime){\rm sinc}\left(\frac{L|\boldsymbol{k}-\boldsymbol{k}^{\prime}|^{2}}{4K}\right),
\end{equation}
where $\boldsymbol{k}$ and $\boldsymbol{k}^{\prime}$ can be used
to denote the wave vectors of the signal and idler photons respectively,
$\eta(\boldsymbol{k})$ is the normalized spectrum function of the
pump beam, $L$ is the length of the nonlinear crystal in the propagating
direction, $K$ is the magnitude of the wave vector of the pump field,
amd ${\rm sinc}(x)=\sin(x)/x$. 

In the following, we focus on twisted photon pairs carrying non-vanishing orbital angular momentum (OAM). Without
loss of generality, we assume the SAF of the two-photon states $\left|P_{\xi}^{\pm}\right\rangle $
to be a simple form, such as
\begin{equation}
\xi^{\pm}(\boldsymbol{k},\boldsymbol{k}^{\prime})=\mathcal{N}\left[\eta_{s}(\boldsymbol{k})\eta_{i}(\boldsymbol{k}^{\prime})e^{i(m_{s}\varphi_{k}+m_{i}\varphi_{k}^{\prime})}\pm\eta_{i}(\boldsymbol{k})\eta_{s}(\boldsymbol{k}^{\prime})e^{i(m_{i}\varphi_{k}+m_{s}\varphi_{k}^{\prime})}\right].\label{eq:xi_entagled}
\end{equation}
Here, $\eta_{s}(\boldsymbol{k})$ and $\eta_{i}(\boldsymbol{k})$
determine the pulse length and pulse shape of the two photons, the
integers $m_{s}$ and $m_{i}$ are the OAM quantum numbers of the two
photons, and $\mathcal{N}$ is a normalization factor. We note that
photon pairs generated from SPDC must satisfy the energy conservation
condition $\omega_{\boldsymbol{k}}+\omega_{\boldsymbol{k}'}=\omega_{p}$ ($\omega_{p}$
the pump frequency). Thus, the two photon are also entangled in frequency
degrees of freedom as shown in Eq.~(\ref{eq:xi_entagled}). However, this spectral entanglement only modifies the HOM interference curve slightly~\cite{branczyk2017hong} and it is not of
much significance in our concerned problem. In experiments, spectral filters can be used to ensure spectral indistinguishability of the two photons and to remove the frequency entanglement~\cite{Scott2017limits,Lyons2018Ashley}.  For degenerate photon pairs generated by a pump without OAM, we have $m_{s}=-m_{i}=m$ and $\eta_{s}(\boldsymbol{k})=\eta_{i}(\boldsymbol{k})=\eta(\boldsymbol{k})$
with normalization constraint $\int d^{3}k\left|\eta(\boldsymbol{k})e^{\pm im\varphi_{k}}\right|^2=1.$
In this case, the SAF $\xi^{\pm}(\boldsymbol{k},\boldsymbol{k}^{\prime})$
reduces to
\begin{equation}
\xi^{\pm}(\boldsymbol{k},\boldsymbol{k}^{\prime})=\mathcal{N}\eta(\boldsymbol{k})\eta(\boldsymbol{k}^{\prime})\left[e^{im(\varphi_k-\varphi_{k^{\prime}})}\pm e^{-im(\varphi_k-\varphi_{k^{\prime}})}\right],\label{eq:xi_entangled-1}
\end{equation}
with $\mathcal{N}=1/\sqrt{2}$. \textbf{We note that the function $\eta(\boldsymbol{k})$
will be assumed to be independent on $\varphi_{k}$ in the following}~\cite{Yang2021}.

The wave-packet function in the real space is given by
\begin{align}
\tilde{\xi}^{\pm}(\boldsymbol{r},\boldsymbol{r}^{\prime},t) & =\mathcal{N}\left[\tilde{\eta}_{m}(\boldsymbol{r},t)\tilde{\eta}_{-m}(\boldsymbol{r}^{\prime},t)e^{im(\varphi-\varphi^{\prime})}\pm\tilde{\eta}_{-m}(\boldsymbol{r},t)\tilde{\eta}_{m}(\boldsymbol{r}^{\prime},t)e^{-im(\varphi-\varphi^{\prime})}\right]\\
 & =\mathcal{N}\tilde{\eta}_{m}(\boldsymbol{r},t)\tilde{\eta}_{m}(\boldsymbol{r}^{\prime},t)\left[e^{im(\varphi-\varphi^{\prime})}\pm e^{-im(\varphi-\varphi^{\prime})}\right],
\end{align}
with
\begin{equation}
\tilde{\eta}_{\pm m}(\boldsymbol{r},t)=\frac{i^{\pm m}}{\sqrt{2\pi}}\int_{-\infty}^{\infty}dk_{z}\int_{0}^{\infty}\rho_{k}d\rho_{k}\eta(k_{z},\rho_{k})J_{\pm m}(\rho\rho_{k})e^{i(k_{z}z-\omega_{\boldsymbol{k}}t)}, \label{eq:eta-tilde}
\end{equation}
where $\rho_k=\sqrt{k^2_x+k^2_y}$. We have used the fact that $J_{m}(x)=(-1)^{m}J_{-m}(x)$ and $\tilde{\eta}_{m}(\boldsymbol{r},t)=\tilde{\eta}_{-m}(\boldsymbol{r},t)$ in
the last step. We note that the function $\tilde{\eta}_{m}(\boldsymbol{r},t)$ is independent on $\varphi$. The normalization constraint in real
space is given by $\int d^{3}r\left|\tilde{\eta}_{\pm m}(\boldsymbol{r},t)\exp\left(\pm im\varphi\right)\right|^{2}=1$.

\section{Quantum description of a beam splitter\label{sec:beamsplitter}}

\begin{figure}
\includegraphics[width=15cm]{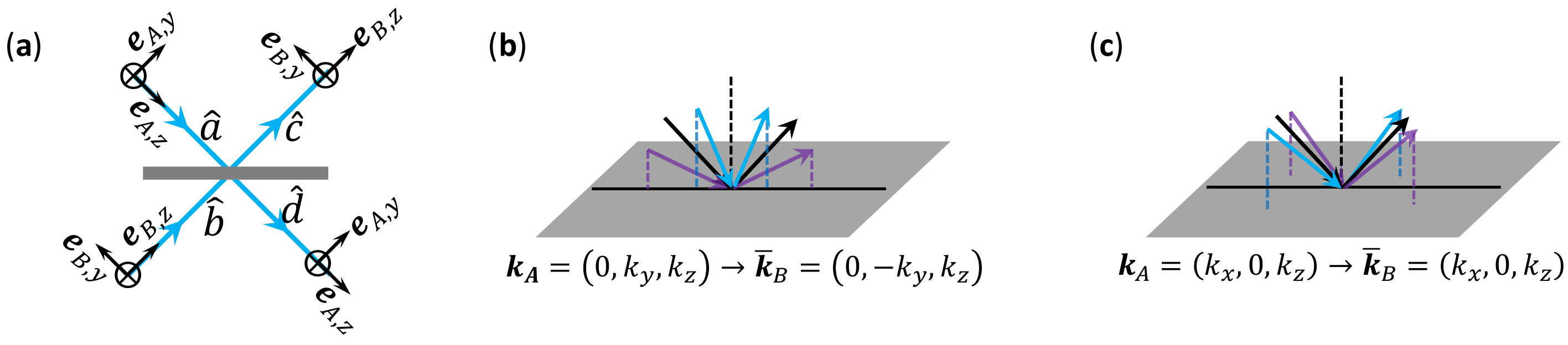}\caption{(a) The transformation between the plane-wave modes and the change of the two coordinate frames at the beam splitter. The modes at the input ports are denoted by the annihilation operators $\hat{a}$ and $\hat{b}$. The modes at the output ports are denoted by annihilation operators $\hat{c}$ and $\hat{d}$. (b) Conversion of a plane-wave mode with a wave vector $\boldsymbol{k}_A=(0,k_y,k_z)$ in the $yz$-plane from frame-$A$ to frame-$B$ after the reflection. The reflection occurs in the $yz$-plane as shown in (a). (c) Conversion of a plane-wave mode with a wave vector $\boldsymbol{k}_A=(k_x,0,k_z)$ in the $xz$-plane from frame-$A$ to frame-$B$  after the reflection. \label{fig:beamsplitter}}
\end{figure}

In the one-dimensional (1D) case, the quantum description of a beam splitter is given by~\cite{loudon2000quantum}
\begin{align}
\hat{c}_{k} & =R_{k}\hat{a}_{k}+T_{k}\hat{b}_{k},\\
\hat{d}_{k} & =T_{k}\hat{a}_{k}+R_{k}\hat{b}_{k},
\end{align}
where $R_k$ and $T_k$ are the reflection and transmission coefficients, respectively. Usually, these two coefficients are approximated as wave-vector-independent constants for a non-dispersive beam splitter.
However, for the three-dimensional (3D) case, the theoretical description of the beam splitter
becomes much more complicated, because the incident angles for different modes could be different. Thus, the reflection and transmission coefficients are usually dependent on $\boldsymbol{k}$. On the other hand, the wave vector $\boldsymbol{k}=(k_x,k_y,k_z)$ of a plane-wave mode in one coordinate frame is converted to $\bar{\boldsymbol{k}}=(k_x,-k_y,k_z)$ in the other frame after a reflection in the two-coordinate-frame formalism as shown in Fig.~\ref{fig:beamsplitter} (b) and (c).

The input-output relations for a beam splitter with 3D incident pulses are given by
\begin{align}
\hat{c}_{\boldsymbol{k}_{B}} & =\left(R_{\boldsymbol{k}}\hat{a}_{\bar{\boldsymbol{k}}_{A}}+T_{\boldsymbol{k}}\hat{b}_{\boldsymbol{k}_{B}}\right),\\
\hat{d}_{\boldsymbol{k}_{A}} & =\left(T_{\boldsymbol{k}}\hat{a}_{\boldsymbol{k}_{A}}+R_{\boldsymbol{k}}\hat{b}_{\bar{\boldsymbol{k}}_{B}}\right).
\end{align}
Here, we see the helical phase factor $\exp(im\varphi_k)=[(k_x+ik_y)/\rho_k]^m$ changes to $\exp(-im\varphi_k)=[(k_x-ik_y)/\rho_k]^m$ in each reflection. Thus, the OAM quantum number of a vortex pulse changes its sign $m\rightarrow-m$ after a reflection. In the paraxial
ray limit, the reflection and transmission coefficients for the $50:50$
beam splitter can still be approximated as $T_{k}=1/\sqrt{2}$ and
$R_{k}=i/\sqrt{2}$. In the following, we will use the inverse transformations
\begin{align}
\hat{a}_{\boldsymbol{k}_{A}} & =\frac{1}{\sqrt{2}}\left(\hat{d}_{\boldsymbol{k}_{A}}-i\hat{c}_{\bar{\boldsymbol{k}}_{B}}\right),\\
\hat{b}_{\boldsymbol{k}_{B}} & =\frac{1}{\sqrt{2}}\left(\hat{c}_{\boldsymbol{k}_{B}}-i\hat{d}_{\bar{\boldsymbol{k}}_{A}}\right).
\end{align}

\section{Hong-Ou-Mandel interference of twisted photon pairs}
Now we consider the HOM interference for an arbitrary input two-photon pulse described by the state in Eq.~(\ref{eq:two-photon-state-1}) (or its equivalent expression (\ref{eq:P_RS}) in the real space). After the beam splitter, the output state is given by
\begin{align}
\left|\Psi_{{\rm out}}\right\rangle = & \frac{1}{2}\int d\boldsymbol{k}\int d\boldsymbol{k}^{\prime}\xi(\boldsymbol{k},\boldsymbol{k}^{\prime})\left[i\hat{d}_{\boldsymbol{k}_{A}}^{\dagger}\hat{d}_{\bar{\boldsymbol{k}}_{A}^{\prime}}^{\dagger}+i\hat{c}_{\bar{\boldsymbol{k}}_{B}}^{\dagger}\hat{c}_{\boldsymbol{k}_{B}^{\prime}}^{\dagger}+\hat{d}_{\boldsymbol{k}_{A}}^{\dagger}\hat{c}_{\boldsymbol{k}_{B}^{\prime}}^{\dagger}-\hat{d}_{\bar{\boldsymbol{k}}_{A}^{\prime}}^{\dagger}\hat{c}_{\bar{\boldsymbol{k}}_{B}}^{\dagger}\right]\left|0\right\rangle, \label{eq:Psi_out1}
\\
= & \frac{1}{2}\int d\boldsymbol{r}\int d\boldsymbol{r}^{\prime}\left[i\tilde{\xi}(\boldsymbol{r},\bar{\boldsymbol{r}}^{\prime},t)\hat{\psi}_{d}^{\dagger}(\boldsymbol{r}_{A})\hat{\psi}_{d}^{\dagger}(\boldsymbol{r}_{A}^{\prime})+i\tilde{\xi}(\bar{\boldsymbol{r}},\boldsymbol{r}^{\prime},t)\hat{\psi}_{c}^{\dagger}(\boldsymbol{r}_{B})\hat{\psi}_{c}^{\dagger}(\boldsymbol{r}_{B}^{\prime})+\tilde{\xi}_{cd}(\boldsymbol{r},\boldsymbol{r}^{\prime},t)\hat{\psi}_{d}^{\dagger}(\boldsymbol{r}_{A})\hat{\psi}_{c}^{\dagger}(\boldsymbol{r}_{B}^{\prime})\right]\left|0\right\rangle,\label{eq:Psi_out2} 
\end{align}
where we have used the relation
\begin{equation}
\hat{d}_{\bar{\boldsymbol{k}}_{A}}=\frac{1}{\sqrt{(2\pi)^{3}}}\int d^{3}r\hat{\psi}(\boldsymbol{r}_{A})e^{-i\bar{\boldsymbol{k}}_{A}\cdot\boldsymbol{r}_{A}}=\frac{1}{\sqrt{(2\pi)^{3}}}\int d^{3}r\hat{\psi}(\boldsymbol{r}_{A})e^{-i\boldsymbol{k}_{A}\cdot\bar{\boldsymbol{r}}_{A}}
\end{equation}
with $\bar{\boldsymbol{r}}=(x,-y,z)$ and
\begin{equation}
\tilde{\xi}_{cd}(\boldsymbol{r},\boldsymbol{r}^{\prime},t)=\tilde{\xi}(\boldsymbol{r},\boldsymbol{r}^{\prime},t)-\tilde{\xi}(\bar{\boldsymbol{r}}^{\prime},\bar{\boldsymbol{r}},t).
\end{equation}
The reversed sign of the $y$ coordinate leads to the well-known result that an image will be reversed by a mirror. The spiral phase in the real space also changes its sign, i.e., $\exp (im\varphi)\rightarrow \exp (-im\varphi)$.

In HOM interference experiments, we usually measure the following quantities: the two-photon-event probability at each output port 
\begin{align}
P_{cc}^{(2)} & =\frac{1}{2}\int d\boldsymbol{r}\int d\boldsymbol{r}'\left\langle \Psi_{\rm out}\right|\hat{\psi}_{c}^{\dagger}(\boldsymbol{r}_{B})\hat{\psi}_{c}^{\dagger}(\boldsymbol{r}_{B}^{\prime})\hat{\psi}_{c}(\boldsymbol{r}_{B}^{\prime})\hat{\psi}_{c}(\boldsymbol{r}_{B})\left|\Psi_{\rm out}\right\rangle = \frac{1}{2}\int d\boldsymbol{k}\int d\boldsymbol{k}'\left\langle \Psi_{\rm out}\right|\hat{c}_{\boldsymbol{k}_{B}}^{\dagger}\hat{c}_{\boldsymbol{k}_{B}^{\prime}}^{\dagger}\hat{c}_{\boldsymbol{k}_{B}^{\prime}}\hat{c}_{\boldsymbol{k}_{B}}\left|\Psi_{\rm out}\right\rangle,\\  
P_{dd}^{(2)} & =\frac{1}{2}\int d\boldsymbol{r}\int d\boldsymbol{r}'\left\langle \Psi_{\rm out}\right|\hat{\psi}_{d}^{\dagger}(\boldsymbol{r}_{A})\hat{\psi}_{d}^{\dagger}(\boldsymbol{r}_{A}^{\prime})\hat{\psi}_{d}(\boldsymbol{r}_{A}^{\prime})\hat{\psi}_{d}(\boldsymbol{r}_{A})\left|\Psi_{\rm out}\right\rangle = \frac{1}{2}\int d\boldsymbol{k}\int d\boldsymbol{k}'\left\langle \Psi_{\rm out}\right|\hat{d}_{\boldsymbol{k}_{A}}^{\dagger}\hat{d}_{\boldsymbol{k}_{A}^{\prime}}^{\dagger}\hat{d}_{\boldsymbol{k}_{A}^{\prime}}\hat{d}_{\boldsymbol{k}_{A}}\left|\Psi_{\rm out}\right\rangle, 
\end{align}
and the coincident probability of the two output ports
\begin{equation}
P_{{cd}}^{(2)} = \int d\boldsymbol{r}\int d\boldsymbol{r}'\left\langle \Psi_{\rm out}\right|\hat{\psi}_{d}^{\dagger}(\boldsymbol{r}_{A})\hat{\psi}_{c}^{\dagger}(\boldsymbol{r}^{\prime}_{B})\hat{\psi}_{c}(\boldsymbol{r}_{B})^{\prime}\hat{\psi}_{d}(\boldsymbol{r}_{A})\left|\Psi_{\rm out}\right\rangle = \int d\boldsymbol{k}\int d\boldsymbol{k}'\left\langle \Psi_{\rm out}\right|\hat{d}_{\boldsymbol{k}_{A}}^{\dagger}\hat{c}_{\boldsymbol{k}_{B}^{\prime}}^{\dagger}\hat{c}_{\boldsymbol{k}_{B}^{\prime}}\hat{d}_{\boldsymbol{k}_{A}}\left|\Psi_{\rm out}\right\rangle.   
\end{equation}

For a photon pair described by Eq.~(\ref{eq:two-photon-state-1}) as the input state of a $50:50$ beam splitter, we have
\begin{equation}
P_{cc}^{(2)} =  \frac{1}{4}+\frac{1}{8}\int d\boldsymbol{k}\int d\boldsymbol{k}'\left[\xi^{*}(\bar{\boldsymbol{k}},\boldsymbol{k}')\xi(\bar{\boldsymbol{k}}^{\prime},\boldsymbol{k})+\xi^{*}(\bar{\boldsymbol{k}}^{\prime},\boldsymbol{k})\xi(\bar{\boldsymbol{k}},\boldsymbol{k}')\right],
\end{equation}
\begin{equation}
 P_{dd}^{(2)}    =\frac{1}{4}+\frac{1}{8}\int d\boldsymbol{k}\int d\boldsymbol{k}'\left[\xi^{*}(\boldsymbol{k},\bar{\boldsymbol{k}}^{\prime})\xi(\boldsymbol{k}^{\prime},\bar{\boldsymbol{k}})+\xi^{*}(\boldsymbol{k}^{\prime},\bar{\boldsymbol{k}})\xi(\boldsymbol{k},\bar{\boldsymbol{k}}^{\prime})\right],
\end{equation}
and
\begin{equation}
P_{{\rm cd}}^{(2)} =    \frac{1}{2}-\frac{1}{4}\int d\boldsymbol{k}\int d\boldsymbol{k}'\left[\xi^{*}(\boldsymbol{k},\boldsymbol{k}')\xi(\bar{\boldsymbol{k}}^{\prime},\bar{\boldsymbol{k}})+\xi^{*}(\bar{\boldsymbol{k}}^{\prime},\bar{\boldsymbol{k}})\xi(\boldsymbol{k},\boldsymbol{k}')\right].
\end{equation}
Next, we apply these results to different input twisted photon pairs.

\textbf{(i) Photon pairs in a product state}:
We first consider a simple case, in which the input photon pair is in a  product state with SAF
\begin{equation}
\xi(\boldsymbol{k},\boldsymbol{k}^{\prime})=\mathcal{N}\xi(\boldsymbol{k})\xi(\boldsymbol{k}^{\prime}).\label{eq:xi_PS}
\end{equation}
Here the normalization constraint is given by $\int d\boldsymbol{k}|\xi(\boldsymbol{k})|^{2}=1$ and $\mathcal{N}=1$. We note that two photons described by the SAF in Eq.~(\ref{eq:xi_PS}) have been perfectly aligned at the beam splitter and their optical path difference has also been carefully matched.

The two-photon-event probability in each output port is given by
\begin{align}
P_{cc}^{(2)} & =\frac{1}{4}\left[1+\int d\boldsymbol{k}\xi^{*}(\bar{\boldsymbol{k}})\xi(\boldsymbol{k})\int d\boldsymbol{k}'\xi^{*}(\boldsymbol{k}')\xi(\bar{\boldsymbol{k}'})\right],\\
P_{dd}^{(2)} & =\frac{1}{4}\left[1+\int d\boldsymbol{k}\xi^{*}(\bar{\boldsymbol{k}})\xi(\boldsymbol{k})\int d\boldsymbol{k}'\xi^{*}(\boldsymbol{k}')\xi(\bar{\boldsymbol{k}'})\right].
\end{align}
The coincidence probability is given by
\begin{equation}
P_{cd}^{(2)} =\frac{1}{2}-\frac{1}{4}\int d\boldsymbol{k}\int d\boldsymbol{k}'\left[\xi^{*}(\boldsymbol{k})\xi(\bar{\boldsymbol{k}})\xi^{*}(\bar{\boldsymbol{k}}^{\prime})\xi(\boldsymbol{k}^{\prime})+\xi^{*}(\bar{\boldsymbol{k}})\xi(\boldsymbol{k})\xi^{*}(\boldsymbol{k}^{\prime})\xi(\bar{\boldsymbol{k}}^{\prime})\right].
\end{equation}
If the SAF of the input pulse has the symmetry $\xi(\bar{\boldsymbol{k}})=\xi(\boldsymbol{k})$ (e.g., the fundamental Gaussian mode~\cite{enderlein2004unified}),
then we will obtain the perfect normal HOM interference with $P_{cd}^{(2)}=0$ and $P_{ cc}^{(2)}=P_{dd}^{(2)}=1/2$. A HOM interference dip can be obtained by adding a delay phase factor $\exp (-i\omega_k \tau)$ to one of the input port~\cite{branczyk2017hong}.

Now we consider a twisted photon pair in a product state with SAF
\begin{equation}
\xi(\boldsymbol{k},\boldsymbol{k}^{\prime})=\mathcal{N}\eta(\boldsymbol{k})\eta(\boldsymbol{k}^{\prime})\exp[im(\varphi_{k}- \varphi_{k}^{\prime})],
\end{equation}
where the two photons carry an equal amount of OAM with the opposite sign. Using the fact that $\eta(\boldsymbol{k})=\eta(k_{z},\rho_{k})$
is independent on $\varphi_{k}$, we have $P_{cc}^{(2)}=P_{dd}^{(2)}=1/2$ and $P_{cd}^{(2)}=0$. If the two photons carry the same amount of OAM,
\begin{equation}
\xi(\boldsymbol{k},\boldsymbol{k}^{\prime})=\mathcal{N}\eta(\boldsymbol{k})\eta(\boldsymbol{k}^{\prime})\exp[im(\varphi_{k}+ \varphi_{k}^{\prime})],
\end{equation}
we have $P_{cc}^{(2)}=P_{dd}^{(2)}=(1+\delta_{m,0})/4$ and
$P_{cd}^{(2)}=(1-\delta_{m,0})/2$. This is similar to the HOM interference for two photons with orthogonal polarizations (distinguishable photons)~\cite{branczyk2017hong}. 

\textbf{(ii) Entangled two photons with opposite OAM quantum number}:
We consider the HOM interference of symmetrically and anti-symmetrically entangled twisted photon pairs described by the SAF in Eq.~(\ref{eq:xi_entangled-1})
\begin{equation}
\xi^{\pm}(\boldsymbol{k},\boldsymbol{k}^{\prime})=\mathcal{N}\eta(\boldsymbol{k})\eta(\boldsymbol{k}^{\prime})\left[e^{im(\varphi_{k}-\varphi_{k}^{\prime})}\pm e^{-im(\varphi_{k}-\varphi_{k}^{\prime})}\right].
\end{equation}
This type of entangled twisted photon pair has previously been denoted as $|\Psi_m^{\pm}\rangle=(|m,-m\rangle\pm |-m,m\rangle)/\sqrt{2}$ for short~\cite{zhang2016engineering}. Our evaluations give the three probabilities $P_{cc}^{(2)}=P_{dd}^{(2)}=1/2$
and $P_{cd}^{(2)}=0$. No coincidence
events will be observed in experiments if the two optical paths have been perfectly matched. A normal HOM dip will be obtained by varying the optical path in one of the input ports.

\textbf{(iii) Entangled two photons with the same OAM quantum number:}
We consider two entangled twisted photons with the same OAM quantum number as the input. The corresponding SAF is given by
\begin{equation}
\xi^{\pm}(\boldsymbol{k},\boldsymbol{k}^{\prime})=\mathcal{N}\eta(\boldsymbol{k})\eta(\boldsymbol{k}^{\prime})\left[e^{im(\varphi_{k}+\varphi_{k}^{\prime})}\pm e^{-im(\varphi_{k}+\varphi_{k}^{\prime})}\right],\label{eq:xiprime}
\end{equation}
with $\mathcal{N}=1/\sqrt{2}$. This type of entangled twisted photon pair has previously been denoted as $|\Phi_m^{\pm}\rangle=(|m,m\rangle\pm |-m,-m\rangle)/\sqrt{2}$ in Ref.~\cite{zhang2016engineering}. For the input state $|\Phi^{+}_m\rangle$, we have $P_{cc}^{(2)}=P_{dd}^{(2)}=1/2$
and $P_{cd}^{(2)}=0$. Thus a normal HOM dip will be observed. For the input state $|\Phi^{-}_m\rangle$, we have $P_{cc}^{(2)}=P_{dd}^{(2)}=0$
and $P_{cd}^{(2)}=1$. Thus a HOM peak will be observed~\cite{zhang2016engineering}. We note that some special care should be paid to the number of reflections in each optical channel. 

\section{Quantum imaging based on the HOM interference}
In this section, we show how quantum imaging with HOM interference works. In our quantum imaging method, the photon in path-$A$ is the imaging photon and the one in path-$B$ is the reference photon. The target object reflects the imaging photon once and imprints its texture information onto the phase factor $\exp[i\Phi(\rho,\varphi)]$ of the imaging photon. The texture information can be extracted via HOM interference as shown in the following.

It is more convenient to handle the quantum imaging process via our formalism in real space. The photon pair at the input ports of the HOM interferometer is described by the state in Eq.~(\ref{eq:P_RS}) and the output state is given by Eq.~(\ref{eq:Psi_out2}). The phase factor $\exp[i\Phi(\rho,\varphi)]$ is contained in the wave-packet function $\tilde{\xi}(\boldsymbol{r},\boldsymbol{r}^{\prime},t)$ as shown in the following. In experiments, we can directly measure the photon number density at each output 
\begin{align}
 n_{c}(\boldsymbol{r},t) = & \left\langle \Psi_{{\rm out}}\right|\hat{\psi}_{c}^{\dagger}(\boldsymbol{r}_{B})\hat{\psi}_{c}(\boldsymbol{r}_{B})\left|\Psi_{{\rm out}}\right\rangle, \\
n_{d}(\boldsymbol{r},t) = & \left\langle \Psi_{{\rm out}}\right|\hat{\psi}_{d}^{\dagger}(\boldsymbol{r}_{A})\hat{\psi}_{d}(\boldsymbol{r}_{A})\left|\Psi_{{\rm out}}\right\rangle,
\end{align}
via a single-photon-detector array, a CCD camera, or other highly sensitive cameras. The correlation function
\begin{equation}
G_{cd}^{(2)}(\boldsymbol{r},\boldsymbol{r}',t) =\left\langle \Psi_{{\rm out}}\right|\hat{\psi}_{d}^{\dagger}(\boldsymbol{r}_{A})\hat{\psi}_{c}^{\dagger}(\boldsymbol{r}_{B}^{\prime})\hat{\psi}_{c}(\boldsymbol{r}_{B}^{\prime})\hat{\psi}_{d}(\boldsymbol{r}_{A})\left|\Psi_{{\rm out}}\right\rangle,
\end{equation}
can also be measured via coincidence counting.

To simplify the coincidence process and enhance the coincidence signal, we can measure the following signals
\begin{align}
\langle\hat{\mathcal{C}}_{c}(\boldsymbol{r},t)\rangle & =\int d\boldsymbol{r}'\left\langle \Psi_{{\rm out}}\right|\hat{\psi}_{c}^{\dagger}(\boldsymbol{r}_{B})\hat{\psi}_{d}^{\dagger}(\boldsymbol{r}_{A}^{\prime})\hat{\psi}_{d}(\boldsymbol{r}_{A}^{\prime})\hat{\psi}_{c}(\boldsymbol{r}_{B})\left|\Psi_{{\rm out}}\right\rangle,\\
\langle\hat{\mathcal{C}}_{d}(\boldsymbol{r},t)\rangle & =\int d\boldsymbol{r}'\left\langle \Psi_{{\rm out}}\right|\hat{\psi}_{d}^{\dagger}(\boldsymbol{r}_{A})\hat{\psi}_{c}^{\dagger}(\boldsymbol{r}_{B}^{\prime})\hat{\psi}_{c}(\boldsymbol{r}_{B}^{\prime})\hat{\psi}_{d}(\boldsymbol{r}_{A})\left|\Psi_{{\rm out}}\right\rangle,
\end{align}
by replacing the detector array at one of the output port with a bucket detector. From the output state in Eq.~(\ref{eq:Psi_out2}), we have
\begin{align}
\langle\hat{\mathcal{C}}_{c}(\boldsymbol{r},t)\rangle & = \frac{1}{4}\int d\boldsymbol{r}'\tilde{\xi}^{*}_{cd}(\boldsymbol{r}',\boldsymbol{r},t)\tilde{\xi}_{cd}(\boldsymbol{r}',\boldsymbol{r},t),\\
\langle\hat{\mathcal{C}}_{d}(\boldsymbol{r},t)\rangle & =\frac{1}{4}\int d\boldsymbol{r}'\tilde{\xi}^{*}_{cd}(\boldsymbol{r},\boldsymbol{r}',t)\tilde{\xi}_{cd}(\boldsymbol{r},\boldsymbol{r}',t).
\end{align}
In following, we focus more on $\langle\hat{\mathcal{C}}_{d}(\boldsymbol{r},t)\rangle$.

Our quantum imaging method is mainly based on the coincidence signal $\langle\mathcal{C}_{d}(\boldsymbol{r},t)\rangle$. Now we define the corresponding signal-to-noise ratio (SNR). We note that in practice we can never measure the ideal density at a single point. Instead, we measure the accumulated signal at a finite small volume $\Delta V$ determined by the pixel area and measuring time. Thus, the measured signal for the pixel labeled by $\boldsymbol{X}$ will be
\begin{align}
\langle\hat{\mathcal{C}}_{d}(X)\rangle & = \int_{\Delta V_X} d\boldsymbol{r}\int d\boldsymbol{r}'\left\langle \Psi_{{\rm out}}\right|\hat{\psi}_{d}^{\dagger}(\boldsymbol{r}_{A})\hat{\psi}_{c}^{\dagger}(\boldsymbol{r}_{B}^{\prime})\hat{\psi}_{c}(\boldsymbol{r}_{B}^{\prime})\hat{\psi}_{d}(\boldsymbol{r}_{A})\left|\Psi_{{\rm out}}\right\rangle
=\frac{1}{4}\int_{\Delta V_X} d\boldsymbol{r}\int d\boldsymbol{r}'\tilde{\xi}^{*}_{cd}(\boldsymbol{r},\boldsymbol{r}')\tilde{\xi}_{cd}(\boldsymbol{r},\boldsymbol{r}'),
\end{align}
where the integral over $\boldsymbol{r}$ is limited in the small volume $\Delta V_X$ determined by the pixel-$\boldsymbol{X}$.
The SNR for the pixel-$\boldsymbol{X}$ under $N$ independent measurements is defined as
\begin{align}
{\rm SNR}_{\rm HOM} (\boldsymbol{X}) & =\frac{\sqrt{N}\langle\hat{\mathcal{C}}_d(\boldsymbol{X})\rangle}{\sqrt{\langle\Delta\hat{\mathcal{C}}^2_d(\boldsymbol{X})\rangle}}=\frac{\sqrt{N}\langle\hat{\mathcal{C}}_d(\boldsymbol{X})\rangle}{\sqrt{\langle\hat{\mathcal{C}}_{d}(\boldsymbol{X})\rangle- \langle\hat{\mathcal{C}}_{d}(\boldsymbol{X})\rangle^2}}.
\end{align}
where we have used the fact that $ \left\langle \Psi_{{\rm out}}\right|\left[\hat{\mathcal{C}}_{d}(\boldsymbol{X})\right]^2\left|\Psi_{{\rm out}}\right\rangle = \langle\hat{\mathcal{C}}_{d}(\boldsymbol{X})\rangle$. Next, we apply our quantum imaging approach to specific twisted photon pairs.

\subsection{Example for quantum imaging with twisted photon pairs~\label{sec:imaging-product-state}}
In this section, we show quantum imaging with twisted photon pairs in a product state, which is described by the wave-packet function
\begin{equation}
\tilde{\xi}(\boldsymbol{r},\boldsymbol{r}^{\prime},t)=\mathcal{N}\tilde{\eta}_{m}(\boldsymbol{r},t)\tilde{\eta}_{m}(\boldsymbol{r}^{\prime},t)e^{i[m(\varphi-\varphi^{\prime})+\Phi(\rho,\varphi)]},
\end{equation}
with $\mathcal{N}=1$. As pointed out previously, the extra phase $\Phi(\rho,\varphi)$ is due
to the texture of the object in channel $A$. In this case, we have
\begin{align}
n_{c}(\boldsymbol{r},t) = & \frac{1}{4}\left|\tilde{\eta}_{m}(\boldsymbol{r},t)\right|^{2}\left\{ 4+\left[(I_{2}-I_{1})e^{i\Phi(\rho,-\varphi)}+{\rm c.c.}\right]\right\}, \\
n_{d}(\boldsymbol{r},t) = & \frac{1}{4}\left|\tilde{\eta}_{m}(\boldsymbol{r},t)\right|^{2}\left\{ 4+\left[(I_{1}-I_{2})e^{i\Phi(\rho,\varphi)}+{\rm c.c.}\right]\right\},
\end{align}
with
\begin{align}
I_{1} & =\int d\boldsymbol{r}'\left|\tilde{\eta}_{m}(\boldsymbol{r}^{\prime},t)\right|^{2}e^{-i\Phi(\rho^{\prime},\varphi^{\prime})},\\
I_{2} & =\int d\boldsymbol{r}'\left|\tilde{\eta}_{m}(\boldsymbol{r}^{\prime},t)\right|^{2}e^{-i\Phi(\rho^{\prime},-\varphi^{\prime})}.
\end{align}
Since $\tilde{\eta}_m(\boldsymbol{r},t)$ is independent on $\varphi$, we can verify that
\begin{equation}
I_2 = \int_{-\infty}^{\infty} dz'\int_0^{\infty}\rho' d\rho' \left|\tilde{\eta}_{m}(\boldsymbol{r}^{\prime},t)\right|^{2}\int_{-2\pi}^0d\varphi' e^{-i\Phi(\rho^{\prime},\varphi^{\prime})}=I_1.    
\end{equation}
Now, we see that the texture information of the target object cannot be directly extracted from photon number density measured at each output port
\begin{equation}
n_{c}(\boldsymbol{r}',t)   =\left|\tilde{\eta}_{m}(\boldsymbol{r}',t)\right|^{2},\ n_{d}(\boldsymbol{r},t)   =\left|\tilde{\eta}_{m}(\boldsymbol{r},t)\right|^{2}.  
\end{equation}

The texture information can be extracted via the coincidence signal
\begin{equation}
\langle\hat{\mathcal{C}}_d (\boldsymbol{r},t)\rangle =\frac{1}{4}\left|\tilde{\eta}_{m}(\boldsymbol{r},t)\right|^{2}\left[2-I_{2}e^{i\Phi(\rho,\varphi)}-I_{2}^{*}e^{-i\Phi(\rho,\varphi)}\right],
\end{equation}
or 
\begin{equation}
\langle\hat{\mathcal{C}}_c (\boldsymbol{r},t)\rangle =\frac{1}{4}\left|\tilde{\eta}_{m}(\boldsymbol{r},t)\right|^{2}\left[2-I_{1}e^{i\Phi(\rho,-\varphi)}-I_{1}^{*}e^{-i\Phi(\rho,-\varphi)}\right].
\end{equation}
We note that our imaging system has been carefully calibrated such that if $\Phi(\rho,\varphi)=0$ then no coincidence signal will be observed. A normal image and a reversed image will be obtained via $\mathcal{C}_d (\boldsymbol{r},t)$ and $\mathcal{C}_c (\boldsymbol{r},t)$, respectively. Usually, the magnitude of the two integrals is smaller than one, $|I_{1}|=|I_{2}|\leq1$.
To increase the contrast of the image, we can remove the background radial distribution due to the twisted photons themselves by defining
the following re-scaled signal, i.e.,
\begin{equation}
\mathcal{S}_{d}(\boldsymbol{r},t)=\frac{\langle\hat{\mathcal{C}}_d (\boldsymbol{r},t)\rangle-n_{d}(\boldsymbol{r},t)/2}{n_{d}(\boldsymbol{r},t)/2}=-\frac{1}{2}\left[I_{2}e^{i\Phi(\rho,\varphi)}+I_{2}^{*}e^{-i\Phi(\rho,\varphi)}\right].
\end{equation}

\subsection{Numerical simulation}
In the main text, we take a Bessel pulse as an example to show the coincidence images. The SAF of Bessel pulse with Gaussian envelop can be approximated as \cite{Yang2021}
\begin{eqnarray}
\begin{aligned}
\eta(\boldsymbol{k}) =\left(\frac{2 \sigma_z^2}{\pi}\right)^{1/4} \operatorname{exp}[-\sigma_z^2(k_z-k_{z,c})^2]\times\left(\frac{2 \sigma_\rho^2}{\pi k_{\bot,c}^2}\right)^{1/4} \exp[-\sigma_\rho^2(\rho_k-\rho_{k,c})^2].
\label{Bessel2}
\end{aligned}
\end{eqnarray}
The first (second) Gaussian function with center value $k_{z,c}=k_c\cos\theta_c$ ($\rho_{k,c}=k_{c}\sin\theta_c$) describes the envelope of the pulse in the propagating direction (transverse plane). In the real space, the pulse length in $z$-direction is characterized by $\sigma_z$ and the size in the transverse plane is characterized by $\sigma_\rho$.  Here, $\theta_c$ is the polar angle of a Bessel pulse and $k_c$ is determined by the center frequency of the pulse $\omega_c = c|k_c|$.  From Eq.~(\ref{eq:eta-tilde}), we have \begin{equation}
\tilde{\eta}(\boldsymbol{r})\approx i^m\sqrt{\frac{k_c\sin\theta_c}{\pi\sigma_z\sigma_{\rho}}} J_m(\rho k_c\sin\theta_c)\exp\left[-\frac{(ct-z\cos\theta_c)^2}{4\sigma_z^2\cos\theta^2_c}+i(k_{z,c}z-\omega_ct)\right].    
\end{equation}

In the main text, we numerically simulate the imaging with a single-photon avalanche diode (SPAD) image sensor consisting of $50\times 50$ pixels. We take the center wavelength as $\lambda_c = 2\pi/k_c =500$~nm, the size of pulse $\sigma_z = 1000 \lambda_c$ (corresponding to pulse length in time $1.67$~ps) and $\sigma_\rho =1000\lambda_c$.  The pitch size of the SPAD is assumed to be $10\ {\rm \mu m}=20\lambda_c$. We only take the diffraction-free Bessel pulse as an example for theoretical demonstration. To obtain a larger cross-section, we set a very small polar angle $\theta_c = 0.001\pi$. In practice, Laguerre-Gaussian or Bessel-Gaussian modes will be used. The cross-section of the pulse increases when leaving the focal plane. A series of photographic lenses are needed to re-focus the photons on the SPAD array.

\end{appendix}
\end{document}